\documentclass[aps,pra,reprint,superscriptaddress,amsmath,amssymb,floatfix]{revtex4-2}

\usepackage[USenglish]{babel}
\usepackage{lmodern}
\usepackage{microtype}
\usepackage{csquotes}

\usepackage{mathtools}
\usepackage{dsfont}
\usepackage{siunitx}
\usepackage{braket}
\usepackage[acronym]{glossaries}
\usepackage{graphicx}
\graphicspath{{imgs/}}

\usepackage{xcolor}
\usepackage{soul}
\usepackage[unicode, colorlinks=true]{hyperref}

\DeclareMathOperator{\vecop}{vec}
\DeclareMathOperator{\Tr}{Tr}
\newcommand{\bigslant}[2]{{\raisebox{.2em}{$#1$}\left/\raisebox{-.2em}{$#2$}\right.}}

\newacronym{dla}{DLA}{dynamical Lie algebra}
\newacronym{dpp}{DPP}{Disciplined Parametrized Programming}
\newacronym{ftqc}{FTQC}{fault-tolerant quantum computing}
\newacronym{flop}{FLOP}{floating-point operation}
\newacronym{gpu}{GPU}{graphics processing unit}
\newacronym{grape}{GRAPE}{Gradient Ascent Pulse Engineering}
\newacronym{hpc}{HPC}{high performance computing}
\newacronym[longplural=linear matrix inequalities]{lmi}{LMI}{linear matrix inequality}
\newacronym{mqv}{MQV}{Munich Quantum Valley}
\newacronym{nisq}{NISQ}{noisy intermediate-scale quantum computing}
\newacronym{qsl}{QSL}{quantum speed limit}
\newacronym{scs}{SCS}{Splitting Conic Solver}
\newacronym{sdp}{SDP}{semidefinite program}
\newacronym{svd}{SVD}{singular value decomposition}

\begin{document}

\title{Quantum speed limits from symmetries: A numerical analysis}
\author{Santana Lujan}
\email{santana.lujan@dlr.de}
\affiliation{Institute for Space Operations and Astronaut Training, German Aerospace Center (DLR), 82234 We{\ss}ling, Germany}
\affiliation{Department of Physics, Friedrich-Alexander-Universit{\"a}t Erlangen-N{\"u}rnberg, 91058 Erlangen, Germany}
\author{Marco Wiedmann}
\affiliation{Department of Physics, Friedrich-Alexander-Universit{\"a}t Erlangen-N{\"u}rnberg, 91058 Erlangen, Germany}
\author{Daniel Burgarth}
\affiliation{Department of Physics, Friedrich-Alexander-Universit{\"a}t Erlangen-N{\"u}rnberg, 91058 Erlangen, Germany}

\begin{abstract}
Target-specific quantum speed limits (QSLs) bound the minimal time required to realize a given operation.
Recent theoretical work derives these limits directly from the symmetries of the dynamical Lie algebra of the control system and a symmetry-restoring perturbation, bypassing the need to solve the equations of motion.
In practice, however, identifying a suitable symmetry and a perturbation that restores it demands strong physical intuition and offers no guarantee of tightness.
We show that, for any given symmetry, the minimal restoring perturbation is the solution of a convex semidefinite program, which can be solved to global optimality and thus yields the tightest achievable speed limit for that symmetry.
This semidefinite program is one building block of our numerical method, alongside an explicit computation of the space of admissible symmetries and a search over this space.
Since this search is non-convex, it is addressed with gradient-based optimization, black-box heuristics, or the injection of physically motivated candidate symmetries.
We apply the method to three examples: a two-qubit \textsc{cnot} gate, a \textsc{swap} on a spinless-particle chain, and a Dicke state preparation on a qubit ring.
Our results improve on previously known symmetry-based bounds, most notably for the \textsc{swap}, where solving the semidefinite program for the known analytical symmetry tightens the bound by a factor of approximately \(2.5\).
For the Dicke state preparation, however, the bounds decrease with system size, which we trace back to a symmetry-independent perturbation ceiling.
This work provides a reproducible numerical procedure for constructing target-specific quantum speed limits directly from the system Hamiltonians and the target operation.
\end{abstract}

\maketitle

\section{Introduction}

As quantum processors transition from the \gls{nisq} era toward fault tolerance, executing operations as rapidly as possible is essential to maximize gate throughput and mitigate decoherence~\cite{preskillQuantumComputingNISQ2018}.
In quantum control, whenever control fields alone cannot implement a desired operation, the finite interaction strength of the drift Hamiltonian imposes an intrinsic minimal duration, known as the \gls{qsl} or controllability time~\cite{kochQuantumOptimalControl2022}.
Historically rooted in time-energy uncertainty relations, fundamental speed limits were formalized by Mandelstam--Tamm and Margolus--Levitin bounds~\cite{mandelstamUncertaintyRelation1991, margolusMaximumSpeed1998}.
While such \glspl{qsl} provide operation-independent, worst-case bounds, practical interest typically centers on specific operations, such as multiqubit entangling gates or state preparations, where worst-case limits are often loose.

Target-specific operation times are commonly benchmarked from above via numerical pulse optimization such as GRAPE~\cite{khanejaOptimalControl2005} or CRAB~\cite{canevaChoppedRandombasisQuantum2011}, which search for time-optimal control fields, or from the quantum brachistochrone equations for optimal Hamiltonian trajectories~\cite{carliniTimeOptimalQuantumEvolution2006,koikeQuantumBrachistochrone2022}.
However, these approaches require explicitly simulating the time evolution of the system, scaling unfavorably with the Hilbert space dimension.
Lower bounds can be obtained from the geometric frameworks of  \emph{distance to uncontrollability}~\cite{burgarthQuantumDistance2022} and its target-specific generalization, the \emph{distance to unreachability}~\cite{wiedmannQuantumSpeedLimits2026}.
These frameworks establish \glspl{qsl} directly from the Lie-algebraic generators of the system, without solving the Schrödinger equation.
Concretely, given a target operation that is not reachable by the controls alone, a symmetry \(S\) of the control-only algebra which is broken only by the drift term is identified.
The obtained speed limit depends inversely on the spectral norm of the minimal perturbation that restores the symmetry.

In practice, identifying valid symmetries and minimal restoring perturbations remains non-trivial: while Ref.~\cite{burgarthQuantumDistance2022} briefly noted that symmetries can be computed by solving a linear system, previous applications relied on system-specific analytical constructions~\cite{wiedmannQuantumSpeedLimits2026}, offering no guarantee on the minimality of the restoring perturbation.
Since the symmetry-restoring condition is linear in the perturbation, minimizing the spectral norm of the perturbation is a convex problem, which we cast as a \gls{sdp}.
This guarantees that, for any candidate symmetry, the computed perturbation is globally minimal, so that the resulting bound can only be improved by choosing a different symmetry.
Solving the \gls{sdp} serves as the inner optimization of our numerical method, which first computes an explicit basis for the space of admissible symmetries via matrix factorizations and then searches this space for the symmetry yielding the tightest bound.
Although optimizing over the symmetry space is non-convex, having an explicit basis enables structured automated searches while also accommodating physically motivated candidate symmetries.
Altogether, this transforms the calculation of target-specific \glspl{qsl} into a systematic, reproducible procedure requiring only the system Hamiltonians and the target operation as input.

We demonstrate the method on three physical control systems.
First, for a \textsc{cnot} gate on locally controlled qubits, automated search reproduces and marginally improves upon the known symmetry-based analytical bound~\cite{wiedmannQuantumSpeedLimits2026}, validating the numerical implementation.
Second, for the \textsc{swap} operation across a spinless-particle chain, solving the \gls{sdp} for the hand-derived symmetry tightens the previous bound by a constant factor.
However, automated heuristic searches over the computed symmetry space struggle to locate this narrow optimum, demonstrating the importance of physically motivated symmetry candidates.
Finally, for a Dicke state preparation on a globally controlled qubit ring~\cite{muratoriWDickestate2025}, the computed speed limit decreases with system size across all sampled symmetries.
We show that the optimization encounters a symmetry-independent perturbation ceiling governed by the drift Hamiltonian, while leaving open whether symmetries that yield tighter bounds exist in the admissible symmetry space. 

The remainder of the paper is organized as follows: Sec.~\ref{sec:theory} reviews the theoretical framework of symmetry-based speed limits, while Sec.~\ref{sec:methods} details the numerical method and its individual stages.
Section~\ref{sec:applications} presents the application to the three example systems, followed by a discussion of scaling trends and computational bottlenecks in Sec.~\ref{sec:discussion}, and concluding remarks in Sec.~\ref{sec:conclusion}.

\section{Theory}
\label{sec:theory}

A bilinear quantum control system on a Hilbert space \(\mathcal{H}\) of dimension \(d\) (\(d < \infty\)) is defined by a constant, always-on drift term \(H_d\) and variable controls \(H_j\), such that the unitary propagator \(U(t)\) evolves according to the Schrödinger equation
\begin{equation}
    \frac{\mathrm{d}}{\mathrm{d}t} U(t) = -\frac{\mathrm{i}}{\hbar}H(t) U(t) \text{, } U(0) = \mathds{1},
\end{equation}
with the time-dependent Hamiltonian
\begin{equation}
    H(t) = H_d + \sum_{j=1}^r f_j(t) H_j.
\end{equation}
We allow for the control pulses \(f_j : \mathds{R} \rightarrow \mathds{R}\) to be chosen from any reasonable set of functions (e.g., piecewise constant).
In particular, we impose no \textit{a priori} bounds on the control pulses, meaning that the speed limits we obtain are fundamental and not only arise from experimental constraints.

For any given bilinear quantum control system, one can determine the set of all reachable unitaries
\begin{equation}
    \mathcal{R} = \{V \in \mathrm{U}(d) \mid \exists T \geq 0, \exists f_j: \mathds{R} \rightarrow \mathds{R} \text{ s.t. } U(T) = V\}
\end{equation}
that can be implemented on that control system.
To do that, one computes the so-called \gls{dla} of the control system
\begin{equation}
    \mathfrak{g} = \langle iH_d, iH_1, iH_2, ... iH_r \rangle_\text{Lie},
\end{equation}
i.e., the smallest Lie algebra that contains the drift and all controls.
The reachable set is then dense in the associated Lie group (Theorem 3.2.1 in Ref.~\cite{dalessandroIntroductionQuantum2021})
\begin{equation}
    \bar{\mathcal{R}} = \exp(\mathfrak{g}).
\end{equation}
If the \gls{dla} is compact, the closure is not needed and we get exactly \(\mathcal{R} = \exp(\mathfrak{g})\).

Since we do not impose bounds on the control functions, every target that can be reached purely by the controls can also be reached arbitrarily fast.
Formally, we introduce the control Lie algebra
\begin{equation}
    \mathfrak{h} = \langle iH_1, ..., iH_r \rangle_\text{Lie}
\end{equation}
and the set of unitaries that is reachable purely from the controls \(\mathcal{R}_\text{ctrl} = \exp(\mathfrak{h})\).
\Glspl{qsl} only occur in this setting if \(\mathfrak{h}\) is strictly smaller than \(\mathfrak{g}\) and therefore if it is in particular not universal.
Proper subalgebras of compact Lie algebras are characterized by the enlargement of the set of so-called quadratic symmetries \cite{zeierSymmetryPrinciples2011}.
These are Hermitian matrices \(S^{(2)} \in \mathds{C}^{d^2 \times d^2}\) that commute with the symmetric embeddings of the drift and control Hamiltonians onto \(\mathcal{H} \otimes \mathcal{H}\)
\begin{equation}
    \forall j = 1,...,r \text{: } [S^{(2)}, H_j \otimes \mathds{1} + \mathds{1} \otimes H_j] = 0.
\end{equation}
The set of all quadratic symmetries is also called the second-order commutant \(\mathfrak{h}^{(2)}\) of \(\mathfrak{h}\).
This set is higher dimensional than the commutant \(\mathfrak{h}^{(1)}\) of \(\mathfrak{h}\) (i.e., all Hermitian matrices \(S^{(1)} \in \mathds{C}^{d \times d}\) for which \([S^{(1)}, H_j] = 0\), \(j = 1,...,r\)), and every element in the commutant has a corresponding quadratic symmetry obtained by the same symmetric embedding.
A Lie algebra \(\mathfrak{h}\) is a proper subalgebra of the compact Lie algebra \(\mathfrak{g}\) if and only if \(\operatorname{dim} \mathfrak{h}^{(2)} > \operatorname{dim} \mathfrak{g}^{(2)}\) \cite{zeierSymmetryPrinciples2011}.
We always have trivially \(\operatorname{dim} \mathfrak{g}^{(2)} \geq 2\) as the identity and the permutation of the tensor components are always trivial quadratic symmetries.

\begin{figure}[t]
    \centering
    \includegraphics[width=0.72\linewidth]{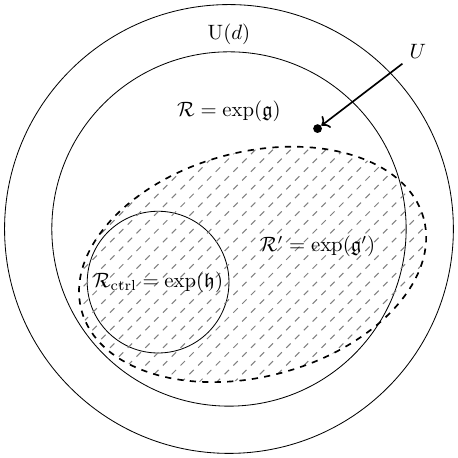}
    \caption{Schematic of the reachable-unitary hierarchy for a controlled system with Hamiltonian \(H = H_d + \sum_j f_j(t) H_j\).
             The system's \acrshort{dla} generates a reachable set \(\mathcal{R} \subseteq \mathrm{U}(d)\) containing the target \(U\), while the controls alone reach a subset \(\mathcal{R}_\text{ctrl} \subset \mathcal{R} \).
             Perturbing the drift restores a symmetry of the control algebra \(\mathfrak{h}\) to the perturbed algebra \(\mathfrak{g}'\), which generates \(\mathcal{R}'\) (dashed ellipse).
             This symmetry forces \(U \notin \mathcal{R}'\), underlying the speed limit for realizing \(U\) of Ref.~\cite{wiedmannQuantumSpeedLimits2026}.
             }
    \label{fig:venn_reachable_operations}
\end{figure}

For any target operation \(U \in \mathcal{R} \setminus \mathcal{R}_\text{ctrl}\) that is not reachable with the control Hamiltonians alone, the constant strength of the drift term imposes a \gls{qsl}, i.e., a minimal time necessary to reach that target.
Since by assumption \(\mathcal{R}\) is strictly larger than \(\mathcal{R}_\text{ctrl}\), we infer that there exists a quadratic symmetry \(S^{(2)} \in \mathfrak{h}^{(2)}\) that does not commute with \(U \otimes U\), i.e., \([U \otimes U, S^{(2)}] \neq 0\).
The guaranteed existence of such a quadratic symmetry is leveraged in Ref.~\cite{wiedmannQuantumSpeedLimits2026} to derive a lower bound on the \gls{qsl} on implementing \(U\)
\begin{equation}
    \label{eq:speed_limit_quadratic}
    T \geq \frac{\lVert [U^{\otimes 2}, S^{(2)}] \rVert_F}{4 \lVert S^{(2)} \rVert_F \lVert \Delta H \rVert_\infty}.
\end{equation}
Here, \(\lVert A \rVert_F = \sqrt{\operatorname{Tr}(A^\dagger A)}\) is the Frobenius norm, \(\lVert \cdot \rVert_\infty\) denotes the spectral norm and \(\Delta H\) is the minimal perturbation to the drift that restores the quadratic symmetry \(S^{(2)}\) to it, i.e.,
\begin{equation}
    \label{eq:restoring_quadratic}
    [S^{(2)}, (H_d + \Delta H) \otimes \mathds{1} + \mathds{1} \otimes (H_d + \Delta H)] = 0.
\end{equation}
In other words, the perturbation \(\Delta H\) should define a (fictional) perturbed control system
\begin{equation}
    \mathfrak{g}' = \langle i(H_d + \Delta H), iH_1, ..., iH_r \rangle_\text{Lie},
\end{equation}
such that \(S^{(2)} \in {\mathfrak{g}'}^{(2)}\).
That the \gls{qsl} depends inversely on the spectral norm of that perturbation was originally shown in Ref.~\cite{burgarthQuantumDistance2022}.
For the relation between all the different Lie algebras and their reachable sets that we introduced in this section, see also Fig.~\ref{fig:venn_reachable_operations}.

If there already exists a linear symmetry \(S^{(1)} \in \mathfrak{h}^{(1)} \setminus \mathfrak{g}^{(1)}\), a tighter bound on the \gls{qsl} can be derived \cite{wiedmannQuantumSpeedLimits2026}
\begin{equation}
    \label{eq:speed_limit_linear}
    T \geq \frac{\lVert [U, S^{(1)}] \rVert_F}{2 \lVert S^{(1)} \rVert_F \lVert \Delta H \rVert_\infty},
\end{equation}
where the perturbation \(\Delta H\) should now restore the linear symmetry instead, i.e.,
\begin{equation}
    \label{eq:restoring_linear}
    [H_d + \Delta H, S^{(1)}] = 0.
\end{equation}
Since Eqs.~\eqref{eq:speed_limit_linear} and~\eqref{eq:speed_limit_quadratic} share the same structure, we combine them into a single bound.
For a symmetry \(S^{(p)}\) of order \(p\), the speed limit is then
\begin{equation}
    \label{eq:speed_limit_combined}
    T \geq \frac{\lVert [U^{\otimes p}, S^{(p)}] \rVert_F}{2 p \lVert S^{(p)} \rVert_F \lVert \Delta H \rVert_\infty}.
\end{equation}
Here, \(p=1\) denotes linear and \(p=2\) quadratic symmetries, and \(\Delta H\) restores \(S^{(p)}\) according to Eq.~\eqref{eq:restoring_linear} or Eq.~\eqref{eq:restoring_quadratic}, respectively.
From here on, we write \(S\) without the superscript whenever a statement holds for both orders.

Finally, notice that any components of \(S^{(2)}\) or \(S^{(1)}\) along \(\mathfrak{g}^{(2)}\) or \(\mathfrak{g}^{(1)}\) do not contribute to the numerators in Eqs. \eqref{eq:speed_limit_quadratic} and \eqref{eq:speed_limit_linear}, but increase the Frobenius norm of the symmetry which appears in the denominator.
Therefore, to get the tightest possible bounds, we should always discard these components, i.e., we should choose the symmetry from the quotient space \(\bigslant{\mathfrak{h}^{(2)}}{\mathfrak{g}^{(2)}}\) or \(\bigslant{\mathfrak{h}^{(1)}}{\mathfrak{g}^{(1)}}\) correspondingly.

\section{Numerical optimization of the speed limit}
\label{sec:methods}

\begin{figure*}[t]
    \centering
    \includegraphics[width=0.75\textwidth]{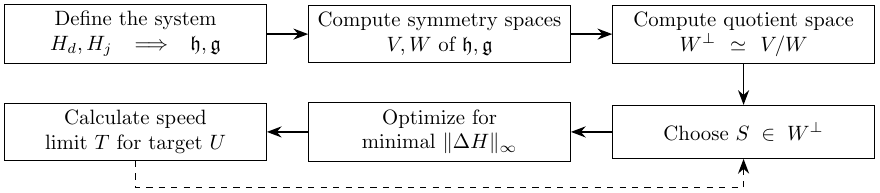}
    \caption{Overview of the numerical procedure for calculating the speed limit \(T\).
    The drift and control Hamiltonians define the dynamical Lie algebras (\acrshortpl{dla}), whose symmetry spaces are computed.
    We calculate a basis for the quotient space via the orthogonal complement, which serves as a pool of candidate symmetries.
    After selecting a symmetry \(S\), solving a semidefinite program (\acrshort{sdp}) yields the minimal-norm perturbation that restores \(S\).
    Together with the target unitary \(U\), the speed limit is calculated via the corresponding linear or quadratic bound.
    The dashed arrow denotes an optional iterative search over \(S\), e.g., using a heuristic or gradient ascent scheme.
    }
    \label{fig:flowchart}
\end{figure*}

This section showcases the mapping of the algebraic framework established in Sec.~\ref{sec:theory} into a systematic numerical procedure, as depicted in Fig.~\ref{fig:flowchart}.
The required input consists solely of a generating set for the Lie algebras \(\mathfrak{h}, \mathfrak{g}\), alongside the target unitary operator \(U\).
Since the speed limit \(T\) represents a lower bound on evolution time, our objective is to maximize \(T\) to obtain the tightest bound.

Following Eq.~\eqref{eq:speed_limit_combined}, we formulate this maximization as a nested optimization problem: we seek a symmetry operator \(S\) from the admissible symmetry space that maximizes the normalized commutator ratio \( \lVert [U^{\otimes p},S] \rVert_F / \lVert S \rVert_F\), while minimizing the spectral norm of the symmetry-restoring perturbation \( \lVert \Delta H \rVert_\infty\).
The method is implemented in Python~3.13 and is structured into three stages:
(1) \emph{Symmetry basis construction}: computing an orthonormal basis for the vector space of symmetries;
(2) \emph{Outer optimization}: determining the optimal symmetry \(S\) within this space; and
(3) \emph{Inner optimization}: solving a convex \gls{sdp} to find the minimal-norm perturbation \(\Delta H\) that restores the symmetry \(S\).

\subsection{Calculating a basis for the vector space of symmetries}
\label{sec:method_symmetry_space}

For a \gls{dla} \(\mathfrak{m}\), we seek all symmetries \(S \in \mathds{C}^{n \times n}\) with \(n = d^p\) that commute with every \(G \in \mathfrak{m}\).
In the following, \( \vecop(\cdot) \colon \mathds{C}^{n \times n} \to \mathds{C}^{n^2} \) denotes row-wise vectorization, mapping a matrix to a vector by stacking its rows.
For quadratic symmetries, we replace each \(G\) by the embedding \(G \otimes \mathds{1} + \mathds{1} \otimes G\).
Using Roth's lemma \( \vecop(ABC) = \left(A \otimes C^T \right) \vecop(B) \), the condition \([G,S] = 0\) is mapped to a linear system
\begin{equation}
    \label{eq:symmetry_linear_equations}
    \vecop(\left[ G,S \right]) = \left( G \otimes \mathds{1} - \mathds{1} \otimes G^T \right) \vecop(S) = 0\,.
\end{equation}

By the Jacobi identity, if an operator \(S\) commutes with two generators \(G_1\) and \(G_2\), it must also commute with their commutator \([G_1, G_2]\), since \([[G_1, G_2], S] = [[G_1, S], G_2] + [G_1, [G_2, S]] = 0\).
Because the \gls{dla} is constructed as the Lie closure of a generating set \(\{G_1, \dots, G_g\}\), enforcing Eq.~\eqref{eq:symmetry_linear_equations} solely on these generators is sufficient to guarantee that \(S\) commutes with the entire \gls{dla}.
Defining the constraint block for each generator as \(M_i \coloneq G_i \otimes \mathds{1} - \mathds{1} \otimes G_i^T \in \mathds{C}^{m \times m}\) with vectorized dimension \(m \coloneq n^2\), we consolidate all individual symmetry requirements into a single joint linear system:
\begin{equation}
    \label{eq:symmetry_space_kernel}
\mathcal{M} \vecop(S) = 0, \quad \text{where} \quad \mathcal{M} \coloneq \begin{pmatrix} M_1 \\ M_2 \\ \vdots \\ M_g \end{pmatrix} \in \mathds{C}^{g m \times m}.
\end{equation}
Here, \(g\) is the number of generators of the \gls{dla}, and the matrix \(\mathcal{M}\) is assembled by vertically stacking the individual blocks \(M_i\) for each generator \(G_i\).

Finding the symmetry space of the \gls{dla} is thus equivalent to computing \(\ker(\mathcal{M})\).
This is a standard null-space problem, which we solve numerically using either the \gls{svd}~\cite[p.~78]{golubMatrixComputations2013} or a rank-revealing QR decomposition~\cite[pp.~276--278]{golubMatrixComputations2013}.
We discuss the computational complexity of this calculation in Sec.~\ref{sec:discussion_limitations_precision_complexity}.
The output of this stage is an explicit orthonormal basis spanning the symmetry space of the \gls{dla}.

Next, we isolate symmetries of the control algebra \(\mathfrak{h}\) not present in the full system algebra \(\mathfrak{g}\).
Let \(V \coloneq \mathfrak{h}^{(p)}\) and \(W \coloneq \mathfrak{g}^{(p)}\) denote the vector spaces of (linear or quadratic) symmetries for the algebras \(\mathfrak{h}\) and \(\mathfrak{g}\), respectively.
Since \(\mathfrak{h} \subset \mathfrak{g}\) by construction, it follows that \(W \subset V\), which allows us to construct the quotient space \(V / W\).
To compute this quotient space numerically, we exploit the fact that in finite-dimensional inner product spaces, \(V / W \simeq W^\perp\)~\cite{provenziEuclideanHilbert2021} for a closed vector subspace \(W\) in \(V\).
The task thus reduces to calculating an orthonormal basis for this orthogonal complement.

To achieve this, let \(\mathbf{V}\) and \(\mathbf{W}\) be matrices whose columns consist of the vectorized orthonormal basis elements of \(V\) and \(W\), respectively.
We perform a thin QR decomposition~\cite[p.~248]{golubMatrixComputations2013} without column pivoting, which preserves the block ordering, of the horizontally concatenated matrix
\begin{equation}
\label{eq:qr_quotient_space}
    \left[\begin{array}{c|c}
        \mathbf{W} & \mathbf{V}
    \end{array}\right]
    \xrightarrow{\ \mathrm{QR}\ }
    \underbrace{
    \left[\begin{array}{c|c|c}
        \mathbf{W} & \mathbf{W^{\perp}} & \ast
    \end{array}\right]
    }_{Q}
    R,
\end{equation}
where  \(Q \in \mathds{C}^{m \times (\dim(W) + \dim(V))}\) has orthonormal columns, and \(R\) is square and upper triangular.
Because the columns of the input matrix are processed sequentially from left to right, the first block of \(Q\) reproduces the basis of \(W\), while the second block \(\mathbf{W^\perp}\) contains exactly the \(k \coloneq \dim(V) - \dim(W)\) columns that span the orthogonal complement \(W^\perp\).
The remaining \(\dim(W)\) columns originate from linearly dependent input columns and are discarded.
Extracting the columns of \(\mathbf{W^\perp}\) yields the vectorized basis elements \(s_j\). By inverting the vectorization, \(S_j \coloneq \vecop^{-1}(s_j)\), we recover a basis of Hermitian operators \(S_j\), enabling us to parametrize any candidate symmetry operator as
\begin{equation}
\label{eq:symmetry_linear_combination}
S = \sum_{j=1}^{k} c_j S_j \quad \text{with} \quad c_j \in \mathds{R}.
\end{equation}

\subsection{Minimal perturbation as a semidefinite program}
\label{sec:method_inner_loop}

Let \(\mathcal{P}(S)\) denote the set of Hermitian perturbations \(\Delta H\) that restore \(S\) according to Eq.~\eqref{eq:restoring_linear} or Eq.~\eqref{eq:restoring_quadratic}. The minimal perturbation norm is then
\begin{equation}
  \label{eq:minimal_perturbation}
  f(S) \coloneq \min_{\Delta H\in\mathcal P(S)} \lVert\Delta H\rVert_\infty .
\end{equation}
Analytical constructions, as in Ref.~\cite{wiedmannQuantumSpeedLimits2026}, provide some \(\Delta H \in \mathcal{P}(S)\), but not necessarily the minimal one.
Since both restoring conditions are linear in \(\Delta H\), Eq.~\eqref{eq:minimal_perturbation} is an instance of spectral-norm minimization under linear constraints.
Such problems are convex and can be cast as an \gls{sdp}~\cite[pp.~169--170]{boydConvexOptimization2023}, so \(f(S)\) can be computed to global optimality.
Introducing an auxiliary variable \(\lambda\) that bounds \(\lVert\Delta H\rVert_\infty\) from above, the \gls{sdp} is posed as:

\begin{equation}
  \label{eq:sdp_formulation}
  \begin{aligned}
    \min_{\lambda\in\mathds{R},\ \Delta H\in\mathcal P(S)} \quad & \lambda \\
    \text{subject to} \quad & \lambda\mathds 1 \pm \Delta H \succeq 0 ,
  \end{aligned}
\end{equation}
with \(f(S)\) as the optimal value \(\lambda^\star\).
The \glspl{lmi} constrain the eigenvalues of \(\Delta H\) to the interval \([-\lambda, \lambda]\), while the membership constraint \( \Delta H\in\mathcal P(S)\) imposes linear constraints, enforcing the physical symmetry.
This formulation also avoids the non-differentiability of \(\lVert\cdot\rVert_\infty\).

To solve Eq.~\eqref{eq:sdp_formulation} numerically, we implement the optimization using the \textsc{cvxpy}~\cite{diamond2016cvxpy} package.
Inside this framework, we declare the perturbation \(\Delta H\) as a Hermitian matrix and \(\lambda\) as a scalar non-negative optimization variable.
To evaluate the inner optimization for different candidate symmetries, we parametrize \(S\) according to Eq.~\eqref{eq:symmetry_linear_combination}, where the Hermitian basis operators \(S_j\) serve as constant matrices and the coefficients \(c_j\) act as parameters within the \gls{dpp} framework~\cite{agrawalDifferentiatingCone2019}.
While held fixed during a single \gls{sdp} evaluation, these parameters are updated dynamically by the outer optimization loop.
This formulation naturally complies with the restrictions of the \gls{dpp} ruleset without requiring workarounds.
Crucially, it allows \textsc{cvxpy} to bypass full problem re-canonicalization at each iteration, reducing the repeated formulation overhead to a small fraction of the initial setup time.

We solve the resulting \gls{sdp} using the MOSEK~\cite{mosekPythonAPI2026} solver for high numerical precision. However, we switch to the \gls{scs}~\cite{donoghueSCSSolver2016}, which supports differentiation through a cone program~\cite{agrawalDifferentiatingCone2019}, when extracting gradient information of the minimal spectral norm with respect to the coefficients \(\nu_j \coloneq \partial f(S) / \partial c_j\).
Further discussion on the choice of solver and numerical precision limits is provided in Sec.~\ref{sec:discussion_limitations_precision_complexity}.

\subsection{Strategies for finding the best symmetry}
\label{sec:method_outer_loop}

The objective of the outer optimization is to maximize the speed limit \(T\) by optimizing the coefficients \(c_j\) in the parametrized symmetry expansion of Eq.~\eqref{eq:symmetry_linear_combination} over a fixed basis \(\{ S_1, \dots,  S_{k} \} \) constructed in Sec.~\ref{sec:method_symmetry_space}.
In contrast to the inner optimization where the symmetry is fixed, allowing these coefficients to vary renders the problem non-convex.
For a fixed symmetry \(S\), the restoring constraint \(\Delta H \in \mathcal{P}(S)\) is strictly linear in the perturbation \(\Delta H\).
However, treating the symmetry as a variable introduces products of the coefficients \(c_j\) and \(\Delta H \) in the commutator of Eq.~\eqref{eq:restoring_linear} or Eq.~\eqref{eq:restoring_quadratic}.
Consequently, the topography of the optimization landscape is dictated by the structure of symmetry space, which itself is determined by the underlying Lie algebras \(\mathfrak{h}\) and \(\mathfrak{g}\).
Because no universal method exists for finding the global maximum in such non-convex landscapes, we employ several complementary strategies to explore the search space.

We first discuss automated methods that operate independently of any \textit{a priori} insight on the underlying physical system.
For systems with low-dimensional symmetry spaces, a grid search over the coefficients \(c_j\) remains computationally feasible, scaling exponentially as \(O(N_s^k)\) for \(N_s\) samples per coefficient dimension across the \(k\)-dimensional space.
This exhaustive mapping of the optimization landscape offers detailed insights into the distribution and shape of extremal areas.
For higher-dimensional symmetry spaces where grid search becomes prohibitive, we employ heuristic and black-box optimization techniques.
As a parallelizable baseline, we sample the coefficients randomly using the \textsc{NumPy}~\cite{harrisArrayProgramming2020} package.
These random samples serve as starting points for standard local optimizers in the \textsc{SciPy} library~\cite{virtanenSciPy102020}, as well as the specialized heuristic solver ZoMBI~\cite{Siemenn2023ZoMBI}, which is tailored particularly for needle-in-a-haystack optimization landscapes.
Finally, we implement a gradient ascent scheme, leveraging the ability of \textsc{cvxpy} to return the perturbation derivative \(\nu_j\) as defined in Sec.~\ref{sec:method_inner_loop}.
Differentiating Eq.~\eqref{eq:speed_limit_combined} with the product rule yields the explicit gradient:
\begin{equation}
\label{eq:speed_limit_gradient}
\begin{aligned}
\frac{\partial T}{\partial c_j} = T \biggl\{
    & \frac{\operatorname{Re}\bigl[ \Tr\bigl( [U^{\otimes p}, S]^\dagger [U^{\otimes p}, S_j] \bigr) \bigr]}{\lVert [U^{\otimes p}, S] \rVert_F^2} \\
    & - \frac{\Tr(S_j S)}{\lVert S \rVert_F^2}
    - \frac{\nu_j}{\lVert \Delta H \rVert_\infty} \biggr\}.
\end{aligned}
\end{equation}

Only the third term in Eq.~\eqref{eq:speed_limit_gradient} depends on the perturbation derivative \(\nu_j\), which is returned by the solver and is therefore only reliable above its convergence tolerance.
To filter out numerical noise, we set \(\nu_j = 0\) whenever the magnitude falls below a threshold \(|\nu_j| \le \eta \coloneq 10^{-4}\), matching the solver's default convergence tolerance.
Following a standard gradient ascent scheme, we update the coefficients at each iteration via \(c_j \to c_j + \epsilon \partial T / \partial c_j\) with a suitable step size \(\epsilon\).
The impact of solver precision and tolerance limits is revisited in Sec.~\ref{sec:solver_choice}.

In contrast to the automated strategies, our method also accepts physically motivated symmetry operators.
This lets us directly use the analytical symmetries constructed in Ref.~\cite{wiedmannQuantumSpeedLimits2026}, allowing physical intuition to guide the search rather than relying solely on automated methods.
Given such a candidate symmetry \(S\), we first verify that it commutes with the control algebra \(\mathfrak{h}\), and optionally project it onto the quotient space \(V / W\).
As discussed in Sec.~\ref{sec:theory}, this discards the components along \(W\), which only increase \(\lVert S \rVert_F\) without contributing to the commutator in Eq.~\eqref{eq:speed_limit_combined}.
The resulting operator is then fed into the inner convex optimization to find the minimal-norm perturbation \(\Delta H\) that restores this symmetry.
The selection of the optimal search strategy depends on the physical system under study, as illustrated in the following section where we apply the method to three distinct physical architectures.

\section{Application to example systems}
\label{sec:applications}

We apply the method of Sec.~\ref{sec:methods} to three physical control systems.
A pair of locally controlled Ising-coupled qubits serves as validation, a spinless particle on a chain shows it can tighten a known analytical bound, and a ring of globally controlled qubits exposes its limitations.

\subsection{CNOT on a pair of locally controlled qubits}
\label{sec:ising_local}

\begin{figure}[tb]
    \centering
    \includegraphics[width=\linewidth]{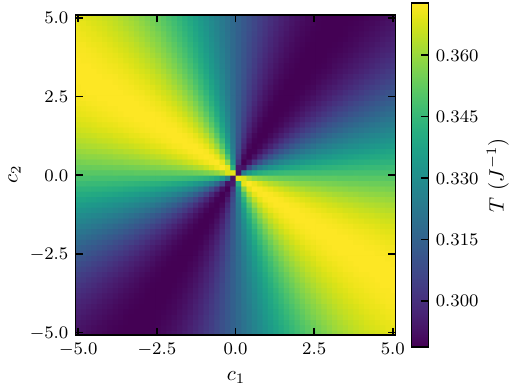}
    \caption{Heatmap of the speed limit \(T\) for the \textsc{cnot} operation between Ising-coupled qubits, obtained from a grid search over the two-dimensional quadratic symmetry \(S^{(2)} = c_1 S^{(2)}_1 + c_2 S^{(2)}_2\) with \(c_1, c_2 \in [-5, 5]\) sampled at 60 points each.
    The inner loop converges to \(\Delta H = -Z_1 Z_2\) with \(\lVert \Delta H \rVert_\infty = 1\) for every sampled \(S^{(2)}\), thus \(T\) is maximized purely by the ratio \( \lVert [U^{\otimes 2},S^{(2)}] \rVert_F / \lVert S^{(2)} \rVert_F\), which peaks along a line through the origin (excluding \(S^{(2)}=0\)) at \(T_\text{max} = 0.373\).}
    \label{fig:cnot_speed_limit}
\end{figure}

We calculate the target-specific speed limit for implementing a \textsc{cnot} gate on a two-qubit system with Hilbert space \(\mathcal{H} = \mathds{C}^2 \otimes \mathds{C}^2\).
The Hamiltonian is composed of a longitudinal Ising coupling \( H_d = JZ_1 Z_2 \) and local controls \(H_1 = X_1\), \(H_2 = Y_1\), \(H_3 = X_2\), \(H_4 = Y_2\) on each qubit.
For our calculations, we set the coupling strength \(J=1\).
Although the controls are purely local and thus incapable of generating entanglement, the system lacks any non-trivial linear symmetry~\cite{wiedmannQuantumSpeedLimits2026}.
Explicit computation of the linear symmetry space, as discussed in Sec.~\ref{sec:method_symmetry_space}, confirms this absence: the null space consists solely of the identity operator, rendering the quotient space \(V / W\) trivial.

Turning to quadratic symmetries, the computed quotient space yields a two-dimensional basis \(\left\{S^{(2)}_1, S^{(2)}_2\right\}\), permitting a thorough, fine-grained grid search over the coefficients \(c_1, c_2\).
We feed each candidate symmetry \(S^{(2)} = c_1 S^{(2)}_1 + c_2 S^{(2)}_2\) directly into our inner optimization \gls{sdp} to compute the corresponding speed limit, with the results shown in Fig.~\ref{fig:cnot_speed_limit}.

For all candidate symmetries, the \gls{sdp} consistently converges to the perturbation \(\Delta H = -Z_1 Z_2\), which fully decouples the qubits and has spectral norm \(\lVert \Delta H \rVert_\infty = 1\).
The speed limit is therefore exclusively dictated by the rate of symmetry breaking on the doubled Hilbert space.
Since the \gls{sdp} always converges to the decoupling perturbation, Eq.~\eqref{eq:speed_limit_quadratic} simplifies to
\begin{equation}
    \label{eq:cnot_simplified_bound}
    T \ge \frac{\lVert [U^{\otimes 2},S^{(2)}] \rVert_F}{4 \lVert S^{(2)} \rVert_F}.
\end{equation}
Searching this parametrized space numerically yields a maximum speed limit of \(T_\text{max} \approx 0.373\) as shown in Fig.~\ref{fig:cnot_speed_limit}.
This result marginally improves upon the analytical bound \(T = \sqrt{2}/4 \approx 0.354\) derived in Ref.~\cite{wiedmannQuantumSpeedLimits2026}, narrowing the gap to the exact solution \(T_\text{\textsc{cnot}} = \pi/4 \approx 0.785\)~\cite{VidalInteractionCost2002}.

The fact that the \gls{sdp} converges to the same perturbation for every sampled symmetry indicates that the symmetry-based bound has saturated for this system.
Nevertheless, this application highlights the advantage of our approach: the ability to determine speed limits directly using structured search methods on an explicit vector space, without ever needing to solve the system's equations of motion.

\subsection{Spinless particle on a chain}
\label{sec:spinless_particle}

\begin{figure}
    \centering
    \includegraphics[width=\linewidth]{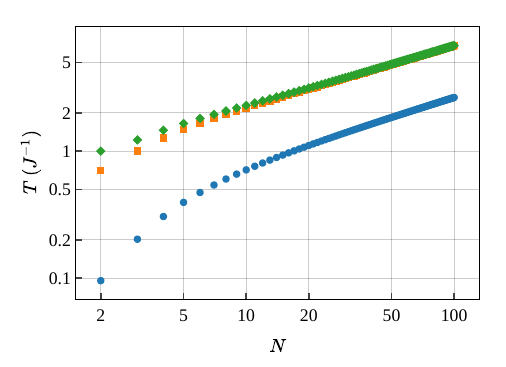}
    \caption{Speed limit \(T\) for the \textsc{swap} operation between two ends of a spinless-particle chain as a function of the number of sites \(N\) (with coupling \(J=1\)).
    The purely analytical bound of Ref.~\cite{wiedmannQuantumSpeedLimits2026} (blue circles, bottom) is compared against two numerically optimized bounds.
    Injecting the analytical symmetry and optimizing the perturbation via the semidefinite program (\acrshort{sdp}) (orange squares, middle) yields an asymptotic tightening by a factor of \(T_\text{num}/T_\text{an} \approx 2.5\).
    Projecting this symmetry onto the numerically constructed symmetry space prior to the optimization (green diamonds, top) further improves the bound at small \(N\) while maintaining the same \(O(\sqrt{N})\) scaling.}
    \label{fig:chain_scaling_speed_limit}
\end{figure}

Next, we consider the case of a single spinless particle on a chain with \(N \in \mathds{N}\) sites.
Adjacent sites are coupled via the drift Hamiltonian
\begin{equation}
    H_d = J \sum_{j=1}^{N-1} \left( \ket{j} \bra{j+1} + \ket{j+1}\bra{j}\right)\,,
\end{equation}
where we control the potential at the first site of the chain
\begin{equation}
    H_1 = \ket{1} \bra{1}.
\end{equation}
For our calculations, we set the coupling strength \(J=1\).
The Hilbert space \(\mathcal{H} = \mathds{C}^N\) grows only linearly with \(N\).
Our target operation swaps the state between two ends of the chain
\begin{equation}
    \text{SWAP} = \mathds{1} - \ket{1} \bra{1} - \ket{N} \bra{N} + \ket{1} \bra{N} + \ket{N} \bra{1}\,,
\end{equation}
which lies within the reachable set of operations, as local control on a boundary node is sufficient for full controllability~\cite{WangSymmetryAndSubs2012,BurgarthZeroForcing2013}.
Our primary focus is the scaling behavior of the speed limit for this operation as a function of chain length \(N\).

We first analyze the symmetry space of the chain.
Unlike in the two-qubit \textsc{cnot} example, this system admits linear symmetries that can be directly used to calculate speed limits.
Since the control Hamiltonian \(H_1\) projects onto the first site, its commutant \(V\) is block-diagonal with dimensions \(1 \times 1\) and \((N-1) \times (N-1)\), yielding \(\dim(V) = 1 + (N-1)^2\).
Full controllability renders the commutant \(W\) of the full \gls{dla} trivial, thus \(\dim(W) = 1\).
Consequently, the quotient space has dimension \(\dim(V/W) = (N-1)^2\).
Due to this quadratic scaling, a thorough search through this optimization space is not feasible for \(N > 3\).

In Ref.~\cite{wiedmannQuantumSpeedLimits2026}, an analytical speed limit for the \textsc{swap} operation was derived by explicitly constructing a single candidate symmetry and a corresponding perturbation, showing an asymptotic scaling of \( O\left( \sqrt{N} \right) \).
A natural starting point is to inject the analytical symmetry candidate directly into our method, allowing the \gls{sdp} solver to optimize the perturbation rather than relying on a hand-constructed one.
The middle trace in Fig.~\ref{fig:chain_scaling_speed_limit} demonstrates that this numerical optimization improves upon the original analytical bound by an asymptotically constant factor.
We extract this factor by fitting the ratio of the bounds, \(R_T = T_\text{num}(N)/T_\text{an}(N)\), to the function \(y(N) = c + a/N + b/N^2\) with \(a,b,c \in \mathds{R}\), which yields \(c = 2.50006 \pm 2.3 \times 10^{-5}\).
The numerical optimization therefore tightens the analytical speed limit by a factor of approximately \(2.5\).
Additionally projecting this candidate symmetry onto the numerically constructed symmetry space prior to solving the \gls{sdp} further tightens the bound for small \(N\), as shown in the topmost trace, while preserving the same \(O\left( \sqrt{N}\right)\) asymptotic scaling.

In the unprojected case, the symmetry \(S^{(1)}_\text{num}\) used as input is identical to the analytical candidate \(S^{(1)}_\text{an}\) for all \(N\).
Consequently, the commutator and symmetry-norm factors in Eq.~\eqref{eq:speed_limit_linear} cancel when taking the ratio \(T_\text{num}/T_\text{an}\), leaving only the dependence on the perturbation norms:
\begin{equation}
    \label{eq:exact_norm_identity}
R_T = \frac{T_\text{num}}{T_\text{an}} = \frac{\lVert \Delta H \rVert_{\infty, \text{an}}}{\lVert \Delta H \rVert_{\infty, \text{num}}} = R_{\lVert \Delta H \rVert_\infty}.
\end{equation}
This isolates the source of our improvement, confirming that it stems entirely from the \gls{sdp} solver finding a perturbation with a significantly smaller spectral norm than the hand-constructed one.
The perturbation designed in Ref.~\cite{wiedmannQuantumSpeedLimits2026} was engineered to introduce a degeneracy in \(H_d\), producing an eigenstate that is also invariant under the control \(H_1\), thus restoring the symmetry.
In contrast, the numerical perturbation merely rotates the existing eigenvectors of \(H_d\) without merging eigenspaces.
Because this rotation is a subtler modification, it requires a smaller spectral norm \(\lVert \Delta H \rVert_\infty\), directly translating into a tighter bound on \(T\).

\begin{figure}
    \centering
    \includegraphics[width=\linewidth]{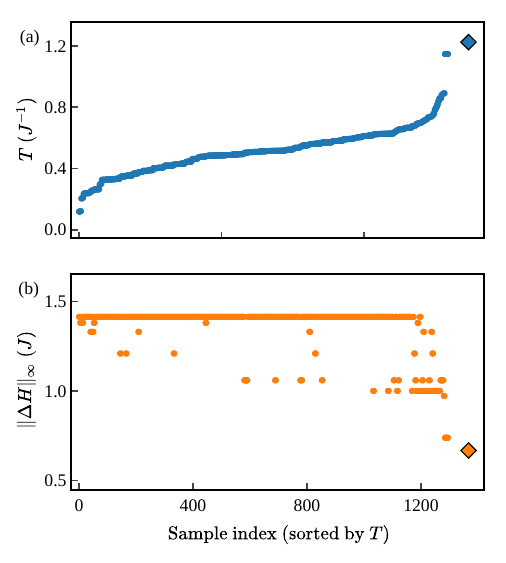}
    \caption{Optimization landscape illustrated by a systematic grid search over the four-dimensional symmetry space for the spinless-particle chain with \(N=3\) sites for \(c_j \in [-1, 1]\) with six samples each, totaling 1296 samples.
    The panels show (a)~the calculated speed limit \(T\) for the \textsc{swap} operation between the ends of the chain and (b)~the corresponding minimal spectral norm of the perturbation \(\lVert \Delta H \rVert_\infty\) obtained by solving the semidefinite program (\acrshort{sdp}), with samples along the horizontal axis sorted by ascending \(T\).
    The best-performing candidate (diamond at the far right) was obtained by directly injecting the physically motivated symmetry rather than from the grid search.
    The vast majority of grid points yield a constant value near \(\lVert \Delta H \rVert_\infty \approx 1.4\), with only rare jumps down to \(\approx 0.74\), whereas the injected symmetry reaches \(\approx 0.67\).
    }
    \label{fig:chain_N_3_flat}
\end{figure}

Motivated by this improvement, we aim to find an even tighter bound by utilizing the automated methods discussed in Sec.~\ref{sec:method_outer_loop}.
Unfortunately, neither heuristic nor black-box methods converge to the known local maximum found with the physically motivated symmetry.
Additionally, due to the high dimensionality for large \(N\), the computational cost for the optimization algorithms scales poorly.
We therefore apply a directional search using the full gradient of Eq.~\eqref{eq:speed_limit_gradient}, in which the \gls{sdp} solver provides the perturbation norm term \(\nu_j\).
We find \(\nu_j\) to oscillate consistently around \(10^{-9}\) across all iterations, falling well below the threshold \(\eta\) established in Sec.~\ref{sec:method_outer_loop} and therefore failing to provide a meaningful search direction.
The remaining two terms depend solely on the symmetry \(S^{(1)}\) and its commutator with the target unitary, thus fully dictating the ascent direction.
As a consequence, the gradients do not steer the optimization toward smaller perturbation norms \(\lVert \Delta H \rVert_\infty\),  yielding speed limits that do not exceed the previous results.
To investigate this behavior, we perform a systematic grid search on the four-dimensional symmetry space of the \(N=3\) chain.
The resulting speed limit \(T\) and perturbation norm \(\lVert \Delta H \rVert_\infty\), plotted in Figs.~\ref{fig:chain_N_3_flat}(a) and \ref{fig:chain_N_3_flat}(b) respectively, reveal an essentially flat \(\lVert \Delta H \rVert_\infty\)-landscape with narrow minima, directly explaining the vanishing gradient components \(\nu_j\).
Resolving the narrow minimum achieved by the physically motivated symmetry requires deliberately matching its coefficients to 8 digits, illustrating why automated search strategies struggle to locate it, even more so in higher dimensions and why physical intuition remains the more effective approach for this system.

\subsection{Ising ring with global control}
\label{sec:ising_global}

As a final example, we apply the method to a ring of \(N \in \mathds{N}\) qubits.
The system evolves under a nearest-neighbor Ising drift Hamiltonian with periodic boundary conditions \(N+1 \equiv 1\)
\begin{equation}
    H_d = J \sum_{n=1}^{N} Z_n Z_{n+1}\,,
\end{equation}
subject to transverse global control fields
\begin{equation}
    H_1 = \sum_{n=1}^{N} X_n\,, \quad H_2 = \sum_{n=1}^{N} Y_n\,.
\end{equation}
We set the coupling strength \(J=1\).
In contrast to the local control setting of Sec.~\ref{sec:ising_local}, these fields act uniformly across the ring, leaving the individual qubits indistinguishable to the controls.
Consequently, the control algebra \(\mathfrak{h}\) is invariant under the full permutation group \(S_N\), whereas
the full algebra \(\mathfrak{g}\) depends on the adjacency structure of the qubits and is therefore symmetric only under the smaller dihedral group \(D_N \subset S_N\)~\cite{wangAcceleratingQuantum2023}.

We evaluate the speed limit for preparing permutation-invariant Dicke states from the all-zero state \(\ket{0} \equiv \ket{0}^{\otimes N}\).
An \(N\)-qubit Dicke state with \(k\) excitations is defined as
\begin{equation}
    \ket{D_k^N} = \binom{N}{k}^{-1/2} \sum_{P} P \left\{ \ket{1}^{\otimes k} \ket{0}^{\otimes N-k} \right\}\,,
\end{equation}
summed over all distinct permutations \(P\) of the site indices~\cite{muratoriWDickestate2025}.
We focus on the two-excitation state, abbreviated as \(\ket{D} \equiv \ket{D_{2}^N}\).
For this target state, numerical pulse optimization indicates that the minimum preparation time scales superlinearly with \(N\)~\cite{muratoriWDickestate2025}, motivating us to examine whether our symmetry-based bounds capture this scaling.

Because \(H_d\), \(H_1\), and \(H_2\) all commute with the action of \(D_N\), the system preserves dihedral symmetry throughout its evolution.
Consequently, no sequence of controls can drive the system outside a given symmetry sector.
Both the initial state \(\ket{0}\) and the target state \(\ket{D}\) are permutation-symmetric and reside entirely within the \(D_N\)-invariant subspace \(\mathcal{H}_\text{inv} \subset \mathcal{H}\).
Following Ref.~\cite{muratoriWDickestate2025}, we construct an isometry \(V_\text{iso}\) whose columns form an orthonormal basis of \(\mathcal{H}_\text{inv}\), reducing the effective Hilbert space dimension from \(2^N\) to \(d = O(2^N / N)\).
We project the Hamiltonians onto this sector via \(H \mapsto V_\text{iso}^\dagger H V_\text{iso}\) for each \(H \in \{H_d, H_1, H_2 \} \), with state vectors transforming as \(\ket{\psi} \mapsto V_\text{iso}^\dagger \ket{\psi}\).
Within this invariant subspace, the state transfer from the initial state to the Dicke state can be realized by the target unitary
\begin{equation}
    \label{eq:ring_target}
    U = \mathds{1} - \ket{0} \bra{0} - \ket{D} \bra{D} + \ket{0} \bra{D} + \ket{D} \bra{0}\,.
\end{equation}
To confirm that \(U \in \mathcal{R}\), we compute \(\mathfrak{g} = \langle iH_d, iH_1, iH_2 \rangle_\text{Lie}\) via nested commutators for ring sizes \(3 \le N \le 6\) and find \(\dim(\mathfrak{g}) = d^2 = \dim(\mathfrak{u}(d))\) in each case.
By the Lie algebra rank condition~\cite{dalessandroIntroductionQuantum2021}, the reachable set is therefore the entire unitary group, \(\mathcal{R} = \mathrm{U}(d)\).

We now turn to the symmetry space of the system, beginning with linear symmetries.
For \(N=3\), no non-trivial solutions exist, and for \(N \ge 4\) all of them commute with the target unitary.
By linearity, the entire space yields only the trivial bound \(T \ge 0 \), forcing us to search for quadratic symmetries.
The dimension of the quotient space grows rapidly with \(N\), yielding \(\dim(V/W)=2, 14, 26, 181\)  for \(N = 3, 4, 5, 6\), respectively.
For systems with \(N \ge 7\), computing this space ceases to be feasible.
We examine the underlying scaling and computational bottlenecks in Sec.~\ref{sec:discussion_limitations_precision_complexity}.

With a numerical basis for the quadratic symmetries at hand, the next step is to determine which symmetry \(S^{(2)}\) yields the tightest speed limit.
Unlike in the case of the spinless-particle chain of Sec.~\ref{sec:spinless_particle}, no analytical candidate is available, prompting us to employ the automated search strategies of Sec.~\ref{sec:method_outer_loop}.
For \(N=3\), the two-dimensional symmetry space allows for an exhaustive grid search.
Across all sampled symmetries \(S^{(2)}\), the minimal perturbation norm remains constant at \(\lVert \Delta H \rVert_\infty = 2\).
When plotted over the coefficient space, the speed limit peaks along a diagonal line through the origin (excluding \(S^{(2)}=0\)), reaching \(T_\text{max} \approx 0.194\).
This profile closely resembles the \textsc{cnot} landscape in Fig.~\ref{fig:cnot_speed_limit}, despite differing in magnitude.
The similarity hints at a shared algebraic structure between the target unitaries, which lies beyond our current scope.

\begin{figure}
    \centering
    \includegraphics[width=\linewidth]{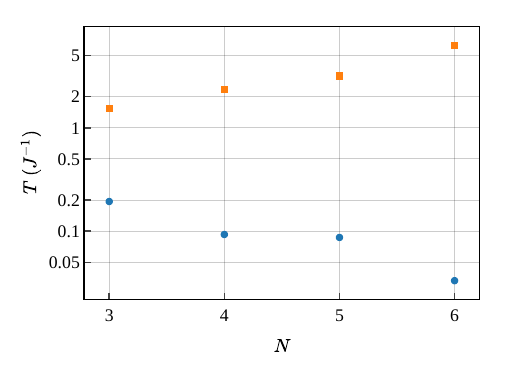}
    \caption{
        Speed limit and numerically optimized preparation times for the Dicke state \(\ket{D_{2}^N}\) in an Ising-coupled qubit ring with global \(X\) and \(Y\) controls as a function of the number of qubits \(N\).
        The symmetry-based lower bound (blue circles) is calculated using the symmetry-based framework.
        The upper trace (orange squares) represents the minimal preparation times obtained via pulse engineering, extracted from Ref.~\cite{muratoriWDickestate2025}, where the \(N=3\) value is inferred from the \(k=1\) state via an instantaneous global \(X\) rotation.
        The symmetry-based lower bound decreases with \(N\), whereas the pulse optimization results exhibit an increasing scaling behavior.}
    \label{fig:ring_speed_limit}
\end{figure}

To extend the analysis to \(N > 3\), we apply the heuristic and random-sampling optimization strategies.
Regardless of the search heuristic or initial distribution, the optimization consistently converges to a constant minimal perturbation norm \(\lVert \Delta H \rVert_\infty \) for each \(N\), matching the observed behavior for \(N=3\).
Furthermore, we observe vanishing gradients \(|\nu_j| < \eta\) across the sampled symmetry space.
Because the optimal perturbation norm remains constant across all sampled symmetries for any given \(N\), it acts as a fixed factor in the denominator of Eq.~\eqref{eq:speed_limit_quadratic}.
Consequently, maximizing the speed limit reduces to maximizing the normalized commutator norm from Eq.~\eqref{eq:cnot_simplified_bound}.
As shown in Fig.~\ref{fig:ring_speed_limit}, our calculated bounds decrease with \(N\).
The bound therefore fails to capture the growth of the preparation time with system size observed in the pulse optimization results of Muratori \textit{et al.}~\cite{muratoriWDickestate2025}.

An inspection of the optimized perturbations \(\Delta H\) shows that their structure depends on the parity of \(N\).
For even \(N\), the solver converges to \(\Delta H = -H_d\), canceling the drift entirely.
For odd \(N\), geometric frustration breaks the spectral symmetry of the ring, guiding the solver to shift the perturbed drift to the identity \(\Delta H = \mathds{1} - H_d\).
Because these perturbed drifts commute trivially with any candidate symmetry, they yield symmetry-agnostic solutions whose spectral properties and consequences for the speed limit scaling are examined in Sec.~\ref{sec:perturbation_ceiling}.

\section{Discussion}
\label{sec:discussion}

In this section, we investigate the mechanism behind the differing scaling behaviors observed across our example systems.
We then examine the computational bottlenecks of our method and discuss our choice of \gls{sdp} solvers.

\subsection{Origin of the differing scaling trends}
\label{sec:perturbation_ceiling}

Across our numerical calculations, the minimal perturbation norm \(f(S)\) defined in Eq.~\eqref{eq:minimal_perturbation} exhibits two distinct convergence behaviors.
For both Ising-coupled qubit systems, \(f(S)\) is completely independent of the candidate symmetry \(S\), remaining constant across all probed symmetries for any given system size \(N\).
In contrast, the spinless particle on a chain yields non-trivial variation in \(f(S)\) as \(S\) varies.
What determines whether \(f(S)\) depends on the choice of symmetry, and how does this invariance determine the scaling behavior of the calculated quantum speed limit \(T\)?

A natural initial hypothesis is to attribute this discrepancy to the symmetry order.
While linear symmetries suffice for the spinless-particle chain, both qubit cases require quadratic symmetries.
Moving from linear to quadratic symmetries affects three properties: it lifts the operators to the doubled Hilbert space \(\mathcal{H} \otimes \mathcal{H}\), algebraically guarantees the existence of non-trivial symmetries for non-fully controllable systems~\cite{zeierSymmetryPrinciples2011}, and introduces an extra factor of \(1/2\) in Eq.~\eqref{eq:speed_limit_quadratic}.
The existence guarantee and the factor of \(1/2\) do not change how \(T\) scales with \(N\).
The lift to \(\mathcal{H} \otimes \mathcal{H}\), however, changes the restoring constraint and may therefore affect the optimal \(\Delta H\).
Since our examples using linear and quadratic symmetries are also different physical systems, we cannot separate the effect of the symmetry order from that of the system.

In both Ising systems, for each \(N\), the \gls{sdp} returns the same perturbation for every sampled symmetry.
This originates from the restoring constraint Eq.~\eqref{eq:restoring_linear} or Eq.~\eqref{eq:restoring_quadratic}, which admits two symmetry-agnostic solutions that map directly to the observed perturbations in Secs.~\ref{sec:ising_local} and~\ref{sec:ising_global}.
The first solution,
\begin{equation}
    \label{eq:pert_cancel}
        \Delta H_\text{cancel} = -H_d\,,
\end{equation}
cancels the drift entirely.
Its spectral norm is given by the spectral radius of the drift Hamiltonian:
\begin{equation}
    \lVert \Delta H_\text{cancel} \rVert_\infty = \max_i \vert \sigma_i \vert\,,
\end{equation}
where \(\sigma_i\) are the eigenvalues of \(H_d\). 
The second construction,
\begin{equation}
    \label{eq:pert_flatten}
        \Delta H_\text{flatten} = c \mathds{1} - H_d\,,
\end{equation}
with \(c \in \mathds{R}\), shifts the drift to a scalar multiple of the identity, collapsing the spectrum into a single degenerate eigenvalue.
This perturbation norm is minimized by centering the shift at the spectral midpoint \(c = (\sigma_\text{max} + \sigma_\text{min})/2\), yielding
\begin{equation}
    \min_c \lVert \Delta H_\text{flatten} \rVert_\infty = \frac{\sigma_\text{max} - \sigma_\text{min}}{2}.
\end{equation}
Because both constructions render the drift equal to zero or a scalar multiple of the identity, the perturbed Hamiltonian and its symmetric embedding commute trivially with any candidate symmetry \(S\).
We define the minimum of these two norms as the symmetry-independent perturbation ceiling:
\begin{equation}
    \label{eq:perturbation_ceiling}
    F(N) \coloneq \min \left(\max_i \vert \sigma_i \vert , \frac{\sigma_\text{max} - \sigma_\text{min}}{2} \right),
\end{equation}
which is determined exclusively by the spectrum of \(H_d\).
This ceiling establishes an upper bound on the convex optimization of Eq.~\eqref{eq:minimal_perturbation}:
\begin{equation}
    \label{eq:perturbation_bound}
    f(S) \le F(N)\,.
\end{equation}

Our numerical optimizations for the Ising-coupled qubits consistently hit this ceiling across the entire sampled space.
By contrast, a grid search on the spinless-particle chain successfully identified symmetries with perturbation norms below this bound, as illustrated in Fig.~\ref{fig:chain_N_3_flat}.
For the qubit examples, even fine grid searches on small systems, like the two-qubit \textsc{cnot} and the three-qubit ring, found no symmetries yielding \(f(S) < F(N)\).
However, because our search relies on finite sampling of the symmetry space, we cannot rule out their existence.

\begin{figure}[t]
    \centering
    \includegraphics[width=3.375in]{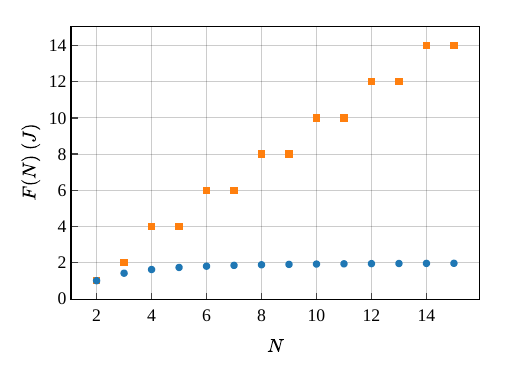}
    \caption{Scaling of the symmetry-independent perturbation ceiling \(F(N)\) with system size \(N\).
    The spinless-particle chain is indicated by the blue circles, while the Ising-coupled qubit ring is indicated by the orange squares.
    The ceiling for the qubit ring grows monotonically, whereas that of the spinless-particle chain asymptotically converges to \(2J\).}
    \label{fig:perturbation_ceil}
\end{figure}

To see how \(F(N)\) dictates the scaling of the speed limit, consider the combined bound of Eq.~\eqref{eq:speed_limit_combined} with \(\lVert \Delta H \rVert_\infty = f(S)\).
Bounding the commutator ratio by applying the triangle inequality and the unitary invariance of the Frobenius norm yields
\begin{equation}
    \label{eq:commutator_bound}
    \frac{\lVert [U^{\otimes p}, S] \rVert_F}{\lVert S \rVert_F} \le \frac{\lVert U^{\otimes p} S \rVert_F + \lVert S U^{\otimes p} \rVert_F}{\lVert S \rVert_F} = 2\,.
\end{equation} 
Since this ratio is bounded by a constant, the asymptotic scaling of the speed limit is dictated entirely by \(1/f(S)\).
By evaluating Eq.~\eqref{eq:perturbation_ceiling} directly from the spectrum of \(H_d\), we obtain the exact ceiling \(F(N)\) independently of the numerical optimization.
As shown in Fig.~\ref{fig:perturbation_ceil}, this ceiling grows linearly with \(N\) for the Ising models, directly accounting for the \(O(1/N)\) decay in the speed limit \(T\).
For the spinless-particle chain, \(F(N)\) grows with \(N\) but rapidly saturates at \(2J\).
Consequently, without symmetries that outperform the trivial solutions, the baseline speed limit would decrease before plateauing to an \(O(1)\) constant.
The achieved \(O\left(\sqrt{N}\right)\) scaling on the chain is therefore not a generic consequence of the perturbation ceiling, but requires finding a symmetry family whose minimal perturbation norm shrinks with \(N\).

The scaling behavior of \(F(N)\) determines the worst-case scaling trend of the speed limit whenever no symmetry yielding \(f(S) < F(N)\) can be found.
Whether the investigated qubit systems are fundamentally constrained by their \glspl{dla} to \(f(S) = F(N)\) across their entire symmetry spaces, or whether favorable symmetries with \(f(S) < F(N)\) exist, remains an open problem.

\subsection{Computational bottlenecks and solver choice}
\label{sec:discussion_limitations_precision_complexity}

\subsubsection{Computational bottlenecks}
The procedure outlined in Fig.~\ref{fig:flowchart} contains two computationally demanding stages: constructing a basis for the symmetry space via the null-space formulation in Eq.~\eqref{eq:symmetry_space_kernel}, and solving the \gls{sdp} in Eq.~\eqref{eq:sdp_formulation} for a given symmetry.
While the symmetry basis is computed upfront and can be stored for subsequent runs, the \gls{sdp} must be solved repeatedly during heuristic and gradient-based searches.
In our calculations, the symmetry basis construction accounts for the primary runtime cost, while initial canonicalization of the \gls{sdp} drives peak memory demand for large systems.

The stacked constraint matrix \(\mathcal{M} \in \mathds{C}^{gm \times m}\) defined in Eq.~\eqref{eq:symmetry_space_kernel} scales rapidly with system size.
Computing its kernel via rank-revealing QR decomposition or \gls{svd} scales as \(O(gm^3)\), with QR carrying a smaller prefactor~\cite[p.~278, pp.~284--285]{golubMatrixComputations2013}.
In terms of the Hilbert space dimension \(d\), this demands \(O(gd^6)\) \glspl{flop} for linear symmetries and \(O(gd^{12})\) for quadratic symmetries.
At larger system sizes, dense representations quickly become prohibitive, necessitating sparse matrix formats and sparse linear algebra routines.

In contrast, the \gls{sdp} formulation does not require vectorization.
The commutator equality constraints are enforced directly on $\mathcal{H}$ for linear symmetries and on $\mathcal{H} \otimes \mathcal{H}$ for quadratic symmetries, while the \glspl{lmi} bounding the spectral norm remain defined on the physical Hilbert space $\mathcal{H}$.
Because the outer optimization queries the \gls{sdp} repeatedly with varying coefficients $c_j$ in Eq.~\eqref{eq:symmetry_linear_combination}, \gls{dpp} significantly reduces the per-iteration overhead.
This benefit is most evident for the largest problem instance considered: the Ising ring example of Sec.~\ref{sec:ising_global} with \(N=6\), whose quotient space \(V / W\) is \(181\)-dimensional.
Initial canonicalization and solving require approximately $13$~minutes, whereas subsequent iterations with updated coefficients drop to roughly $16$~seconds.
During the canonicalization step, memory rather than runtime becomes the limiting factor, with the solver's internal data structures peaking at approximately $94$~GB.

Reducing the Hilbert space dimension $d$ directly addresses both bottlenecks.
As demonstrated in Sec.~\ref{sec:ising_global}, projecting onto invariant subspaces, such as the $D_N$-invariant sector of the ring, reduces $d$ from $2^N$ to $O(2^N/N)$, making larger systems computationally accessible.
However, this approach is applicable only when both the Hamiltonian dynamics and the target unitary leave the chosen subspace invariant, which restricts the class of addressable target gates.

\subsubsection{Solver choice}
\label{sec:solver_choice}
The \textsc{cvxpy} modeling framework~\cite{diamond2016cvxpy} supports multiple backend solvers differing in their underlying algorithms and licensing terms.
We employ both the open-source solver \gls{scs}, \textsc{cvxpy}'s default solver for \glspl{sdp}, which implements Douglas--Rachford splitting~\cite{donoghueSCSSolver2016}, and MOSEK, a commercial interior-point solver available under an academic license~\cite{mosekPythonAPI2026}.
Under default configurations, \gls{scs} terminates at a moderate-precision residual tolerance of \(10^{-4}\), whereas MOSEK reaches a high-precision relative duality gap of \(10^{-8}\).
Furthermore, as described in Sec.~\ref{sec:method_outer_loop}, the \gls{scs} solver supports differentiating the optimal solution with respect to problem parameters, enabling gradient-based optimization.

Our selection of solver is pragmatic: because individual solve times remain modest during our calculations, we default to MOSEK with its higher precision for all heuristic and black-box methods that do not require gradients, switching to \gls{scs} otherwise.
The variation in solver precision does not affect our conclusions, as distinguishing meaningful differences in the optimal perturbations and the resulting speed limit \(T\) does not require a resolution beyond \(10^{-4}\).

\section{Conclusion and Outlook}
\label{sec:conclusion}

In this work, we translated the theoretical framework of target-specific, symmetry-based \glspl{qsl}~\cite{wiedmannQuantumSpeedLimits2026} into a systematic numerical method.
Requiring only the drift and control Hamiltonians alongside the target operation as inputs, the procedure decomposes into three primary stages:
constructing an explicit basis for the admissible symmetry space, searching this space for an optimal candidate symmetry, and optimizing for the minimal-norm perturbation that restores it.
The first and third stages reduce to well-posed computational tasks: a kernel computation and a convex \gls{sdp}, respectively, both guaranteed to yield the exact or globally optimal solution.
In particular, the \gls{sdp} removes the need for hand-constructed perturbations. 
The intermediate stage, which strongly influences the tightness of the resulting speed limit, is non-convex and relies on complementary strategies, including gradient ascent, black-box heuristics, and physically motivated candidate injections.

Applying the method to three physical architectures illustrates both its practical utility and current limitations.
In the two-qubit \textsc{cnot} example, an automated search verified that our numerical optimization can reproduce and slightly tighten the hand-derived symmetry bound of Ref.~\cite{wiedmannQuantumSpeedLimits2026}.
For the \textsc{swap} operation between the ends of a spinless-particle chain, the \gls{sdp} solver discovered a perturbation mechanism based on rotating existing eigenstates of the drift Hamiltonian rather than forcing spectral degeneracy, tightening the analytical bound by a factor of approximately \(2.5\) while preserving the \(O(\sqrt{N})\) scaling.
Conversely, for the Dicke state preparation on a globally controlled qubit ring, the calculated bound decayed as \(O(1/N)\) for all sampled symmetries, failing to capture the superlinear scaling observed via pulse optimization~\cite{muratoriWDickestate2025}.

This behavior originates from the \gls{sdp} solver returning only symmetry-independent perturbations for all sampled candidate symmetries in the qubit architectures.
This pattern revealed a symmetry-independent perturbation ceiling, \(F(N)\), set by perturbations that cause the perturbed drift Hamiltonian to commute trivially with any operator, specifically by canceling the drift entirely or shifting it to a scalar multiple of the identity. 
Because this ceiling depends solely on the spectrum of the drift Hamiltonian, it explains the worst-case scaling behavior of the speed limits observed in the qubit systems.
It also highlights the value of physically motivated candidate symmetries, which proved decisive for achieving tighter bounds in the spinless-particle case.

With this method in place, an immediate next step is to evaluate target-specific \glspl{qsl} across other systems and operations, particularly where bounds are currently loose or unknown.
Beyond these direct applications, several avenues emerge to extend and advance the methodology itself:

\emph{Escaping the perturbation ceiling}: Symmetries that satisfy \(f(S) < F(N)\) appear to be rare and localized within the symmetry space.
Establishing analytical criteria to determine whether a given \gls{dla} admits such symmetries would serve a dual purpose: it would prevent fruitless searches when such solutions are algebraically ruled out, while justifying computational investment in challenging searches when tighter bounds are guaranteed to exist.
Empirically, using the method to investigate the effect of strategically placed local controls in globally driven systems, such as the Ising ring, could expose additional structure in the symmetry space and help bypass the ceiling.

\emph{Numerics-to-theory feedback}: The discovery of a non-obvious perturbation mechanism in the spinless-particle chain demonstrates that convex optimization can find solutions missed by intuition.
Leveraging such solver-discovered perturbations to extend the repertoire of symmetry-perturbation constructions could complement theoretical efforts, closing the loop between the numerical and analytical approaches.

\emph{Algorithmic efficiency}: The primary computational bottleneck remains the null-space calculation of the stacked constraint matrix defining the symmetry space.
An alternative would be to compute pairwise null-space intersections~\cite[pp.~328--329]{golubMatrixComputations2013} sequentially across the generators, restricting each new set of constraints to the current candidate subspace rather than decomposing the entire stacked matrix in a single computation.
This could reduce the computational cost, thus extending the reach to larger systems.

\emph{State-to-state extension}: Extending the theoretical framework beyond unitary and simulable-Hamiltonian targets to state-to-state transfers would broaden the class of problems it can address.
Such an extension would directly apply to the Dicke state preparation studied in this work, which is fundamentally a state-to-state problem currently handled by choosing one specific unitary among the many that implement this transfer.

\begin{acknowledgments}
This research was supported by the Bavarian Ministry of Economic Affairs, Regional Development and Energy with funds from the Hightech Agenda Bayern and is part of the funding measure Munich Quantum Valley.
The authors gratefully acknowledge the computational and data resources provided through the joint high-performance data analytics (HPDA) project “terrabyte” of the German Aerospace Center (DLR) and the Leibniz Supercomputing Center (LRZ).
\end{acknowledgments}

\section*{Data Availability}
The data that support the findings of this article are available from the corresponding author upon reasonable request.

\bibliography{references}

\end{document}